\documentclass[agupp,aguplus]{aguplus}
\usepackage{amssymb}
\usepackage{graphics}
\usepackage{epsfig}

\slugcomment{Submitted to {\it GRL}, 1 November 2010}
\lefthead{MISHIN AND PEDERSEN}
\righthead{HF-INDUCED IONIZATION}
\authoraddr{Space Vehicles Directorate, Air Force Research Laboratory,  Hanscom AFB, MA 01731 (e-mail: Evgeny.Mishin@hanscom.af.mil; Todd.Pedersen@hanscom.af.mi)}
\input{tcilatex}
\begin{document}

\title{Ionizing wave via high-power HF acceleration }
\author{ Evgeny Mishin and Todd Pedersen \\
\noindent Space Vehicles Directorate, Air Force Research Laboratory, Hanscom
AFB, MA, USA}

\begin{abstract}
Recent ionospheric modification experiments with the 3.6 MW transmitter at
the High Frequency Active Auroral Research Program (HAARP) facility in
Alaska led to discovery of artificial ionization descending from the nominal
interaction altitude in the background \textit{F}-region ionosphere by $\sim 
$60 km. This paper presents a physical model of an ionizing wavefront
created by suprathermal electrons accelerated by the HF-excited plasma
turbulence.
\end{abstract}

\section{ Introduction}

High-power {HF }radio waves can excite electrostatic waves in the ionosphere
near altitudes where the injected wave frequency $f_{0}$ matches either the
local plasma frequency $f_{p}\approx 9\sqrt{n_{e}}$ kHz (the density $n_{e}$
in cm$^{-3}$) or the upper hybrid resonance $f_{uhr}=\sqrt{%
f_{p}^{2}+f_{c}^{2}}$ ($f_{c}$ is the electron cyclotron frequency) [e.g., 
\textit{Gurevich}{, 1978]}. The generated waves increase the bulk electron
temperature to $T_{e}=$0.3-0.4 eV, while some electrons are accelerated to
suprathermal energies $\varepsilon =\frac{1}{2}mv^{2}$ up to a few dozen eV
[e.g., \textit{Carlson et al}., 1982; \textit{Rietveld et al.}, 2003]. Upon
impact with neutrals ($N_{2}$, $O_{2}$, $O$), suprathermal electrons excite
optical emissions termed Artificial Aurora (AA) [e.g., \textit{Bernhardt et
al}., 1989; \textit{Gustavsson and Eliasson}, 2008].

Heating-induced plasma density modifications are usually described in terms
of chemical and transport processes [e.g.,\textit{\ Bernhardt et al}., 1989; 
\textit{Djuth et al.}, 1994; \textit{Dhillon and Robinson}, 2005; \textit{%
Ashrafi et al}., 2006]. However, the \textit{Pedersen et al}. [2009; 2010]
discovery of rapidly descending plasma layers seems to point to additional
mechanisms. \textit{Pedersen et al}. [2010, hereafter P10] suggested that
the artificial plasma is able to sustain interaction with the transmitted HF
beam and that the interaction region propagates (downward) as an ionizing
wavefront. In this paper, the formation of such downward-propagating
ionizing front is ascribed to suprathermal electrons accelerated by the
HF-excited plasma turbulence.

\section{Ionizing wave}

\begin{figure}
\noindent\includegraphics[width=20pc]{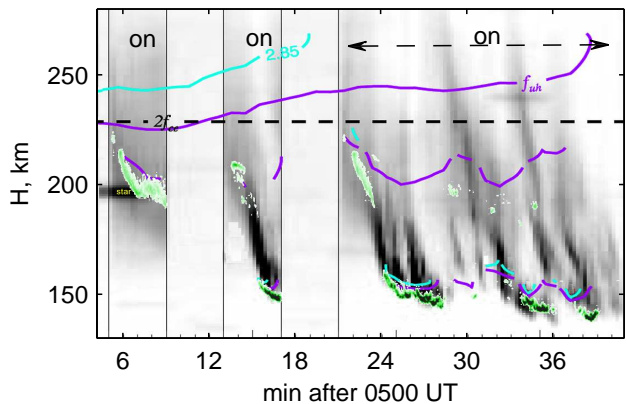}
\caption{  Time-vs-altitude plot of 557.7 nm optical emissions (black color) along 
\textbf{B}$_{0}$. Blue (violet) lines indicate the matching altitudes $f_{0}=
$ $f_{p}$ ($f_{uh}$). The dashed line indicates $h_{2f_{c}}$. The
transmitter on periods are indicated. Shown in green is the MUIR\ IL
intensity (courtesy of Chris Fallen). Horizontal blips are stars passing through the view.}
\label{layer}
\end{figure}

The descending feature is evident in Figure~\ref{layer}, which is
representative of P10 Fig. 3 with the regions of ion-line (IL) radar echoes
from the MUIR incoherent scatter radar (courtesy of Chris Fallen) overlaid
[c.f. \textit{Oyama et al}., 2006]. Shown are sequential altitude profiles
of the green-line emissions ($\lambda =$557.7 nm, excitation potential $%
\varepsilon _{g}\approx $4.2 eV) observed at the HAARP facility on 17 March
2009. Here, the \textit{O}-mode radio beam was injected into the magnetic
zenith (MZ), i.e. along the magnetic field $\mathbf{B}_{0}$, at the
effective radiative power $P_{0}[\mathrm{MW}]\approx $440 and frequency $%
f_{0}=$2.85 MHz (2$f_{c}$ at $h_{2f_{c}}\approx $230 km). The contours of $%
f_{p}=f_{0}$ or $n_{e}=n_{c}\approx 10^{5}$ cm$^{-3}$ (cyan) and $%
f_{uhr}=f_{0}$ (violet) are inferred from ionograms acquired at 1 min
intervals [P10 Fig. 3]. The regions of enhanced IL are shown in green color.
Note, the blue-line emissions at 427.8 nm (not shown) coincided with the
green-line emissions, as seen looking from the HAARP\ site [P10].

During the first 2 min in the heating, the artificial plasma is confined to
the bottomside of the \textit{F} layer at altitudes $h>$180 km. The
corresponding descent of the IL scatter is similar to that described by 
\textit{Dhillon and Robinson} [2005] and \textit{Ashrafi et al}. [2006]. A
sudden brightening of AA and increased speed of descent of the artificial
plasma `layer' (patch) in the HF-beam center occurs near 180 km, while its
peak plasma frequency $foFa$ reaches $f_{0}$. In fact, the optical data
shows [P10] that this patch is fairly uniform near $\sim $180 km but then
becomes a $\sim $20-km\textbf{\ }collection of\textbf{\ }($\parallel $%
\textbf{B}$_{0}$) filaments a few km in diameter. While the degree of
inhomogeneity of the descending patch increases, its speed, $V_{obs}\simeq $%
0.3 km/s, appears to be constant until $\approx $160 km. Then, the
artificial plasma slows down, staying near the terminal altitude $h_{\min
}\approx $150 km before the emissions retreat in altitude near the end of
4-min injection pulse. During a continuous `on' period, the artificial
plasma near $h_{\min }$ was quenched several times, initiating the process
over again from higher altitudes.

Hereafter, we focus on the descending feature at $h\leq $180 km where $%
foFa\geq f_{0}$ or $n_{e}\geq n_{c}$. Enhanced 427.8-nm emissions indicate
the presence of electrons with energies $\varepsilon >\varepsilon
_{b}\approx $18.7 eV, exceeding the ionization energies of $N_{2}$, $O_{2}$,
and $O$. The ionization rate $q_{a}$ is given by 
\begin{equation}
dn_{e}/dt=q_{a}=n_{a}\cdot \left\langle \nu _{ion}(\varepsilon )\right\rangle
\label{dne}
\end{equation}%
where $\nu _{ion}$ is the ionization frequency, and $\left\langle
...\right\rangle $ means averaging over the accelerated distribution of the
density $n_{a}$. Hereafter, we employ \textit{Majeed and Strickland's}
[1997] electron impact cross-sections and the \textit{Hedin} [1991] MSIS90
model for the densities $[N_{2}]$, $[O]$, and $[O_{2}]$ on 17 March 2009.

At each time step $t_{i}$, artificial ionization occurs near the critical
altitude $h_{c}(t_{i})$, defined from the condition $n_{e}(h_{c})=n_{c}$.
The density profile just below $h_{c}$ is represented as follows%
\begin{equation}
n_{e}(x,t_{i})=n_{c}\cdot \Psi \left( x\right)  \label{ne}
\end{equation}%
Here $x=\xi /L_{\parallel }$, $\xi =\left( h_{c}-h\right) /\cos \alpha _{0}$
is the distance along $\mathbf{B}_{0}$, $\alpha _{0}$ is the conjugate of
the magnetic dip angle ($\approx $15$^{\circ }$ at HAARP), and $L_{\parallel
}$ is the ($\parallel $\textbf{B}$_{0}$) extent of the ionization region. $%
\Psi (x)$ is a monotonic function satisfying the conditions $\Psi (0)\geq $1
and $\Psi (x)\ll $1 at $x>$1, since the ambient plasma density $n_{0}\ll
n_{c}$ at $h\leq $180 km. As the ratio $\delta _{e}(\varepsilon )$ of
inelastic ($\nu _{il}$) to elastic ($\nu _{el})$ collision frequencies is
small, the accelerated electrons undergo fast isotropization due to elastic
scattering and thus $L_{\parallel }\simeq \left\langle l_{ion}\sqrt{\delta
_{e}/2}\right\rangle $, where $l_{ion}=v/\nu _{ion}$ [c.f. \textit{Gurevich
et al}., 1985].

Evidently, as soon as at some point $x_{i}\leq 1$ the density $%
n_{e}(x_{i},t_{i}+\Delta t)\simeq q_{a}(\xi _{i})\cdot \Delta t$ reaches $%
n_{c}$, the critical height shifts to this point, i.e. $h_{c}(t_{i+1})\simeq
h_{c}(t_{i})-L_{\parallel }\cdot x$. These conditions define the ionization
time, $T_{ion}^{-1}\simeq q_{a}/n_{c}$, and the speed of descent 
\begin{equation}
V_{d}=\left\vert dh_{c}/dt\right\vert \simeq L_{\parallel
}T_{ion}^{-1}\simeq \left\langle v\sqrt{\delta _{e}/2}\right\rangle
n_{a}/n_{c}  \label{vdesc}
\end{equation}

Note, eq. (\ref{vdesc}) contains no dependence on the total neutral density $%
N_{n}$ and hence predicts $V_{d}\simeq \mathrm{const}(h)$, if the same
distribution of accelerated electrons is created at each step. As $%
\left\langle \delta _{e}^{1/2}v\right\rangle \simeq $1.5$\cdot 10^{6}$ m/s,
we get from eq. (\ref{vdesc}) that the value of $V_{d}$ (\ref{vdesc})
matches $V_{obs}$ at $n_{a}=n_{a}^{(d)}\simeq $6$\cdot 10^{-4}n_{c}$ or $%
n_{a}^{(d)}\simeq $60 cm$^{-3}$.

\section{Discussion and conclusions}

We now turn to justify this acceleration-ionization-descent scenario.
Enhanced IL echoes, like in Figure~\ref{layer}, usually result from the
parametric decay instability (PDI$_{l}$) and oscillating two stream
instability (OTSI) of the pump wave near the plasma resonance [e.g., \textit{%
Mj\o lhus et al}., 2003]. The latter develops if the relative pump wave
energy density $\widetilde{W}_{0}=\left\vert E_{0}^{2}\right\vert /8\pi
n_{c}T_{e}$ exceeds $\widetilde{W}_{th}\simeq \frac{2}{kL_{n}}+\frac{4\nu
_{T}}{\omega _{0}}$, where $k$ is the plasma wave number, $\nu _{T}$ is the
collision frequency of thermal electrons, and $L_{n}^{-1}=\left\vert \nabla
\ln n_{e}\right\vert $. The free space field of the pump wave is $%
E_{fs}\approx $5.5$\sqrt{P_{0}}/r\approx $0.65 V/m at $r=$180 km (at the
HF-beam center) or $\widetilde{W}_{fs}\simeq $5$\cdot 10^{-4}$ at $T_{e}=$%
0.2 eV. For incidence angles $\theta <\arcsin \left( \sqrt{%
f_{c}/(f_{c}+f_{0})}\sin \alpha _{0}\right) $, the amplitude in the first
Airy maximum is $E_{A}\approx (2\pi /\sin \alpha
_{0})^{2/3}(f_{0}L_{n}/c)^{1/6}E_{fs}$ [e.g., \textit{Mj\o lhus et al}.,
2003] or $\widetilde{W}_{A}\approx 0.1(L_{n}/L_{0})^{1/3}$, where $L_{0}=$30
km. For injections at MZ, following \textit{Mj\o lhus et al}. [2003] one
gets $\widetilde{W}_{A}^{(mz)}\approx \widetilde{W}_{A}/4$.

As $\widetilde{W}_{A}^{(mz)}\gg \mu \ $(the electron-to-ion mass ratio), we
get $k\simeq r_{D}^{-1}\left( \mu \widetilde{W}_{A}^{(mz)}\right) ^{1/4}$
[e.g., \textit{Alterkop et al}., 1973] or $kr_{D}\simeq 1/40$ ($r_{D}$ is
the Debye radius) yielding $\widetilde{W}_{th}\simeq 10^{-4}(L_{0}/L_{n})$.
The `instant' gradient-scale of the artificial layer is $L_{n}\simeq $3$%
\rightarrow $1 km at 180$\rightarrow $150 km (see below) gives $\widetilde{W}%
_{th}\approx $(1$\rightarrow $3)$\cdot 10^{-3}$. Thus, OTSI can easily
develop in the first Airy maximum. In turn, PDI$_{l}$ can develop in as many
as $\simeq $30 Airy maxima over a distance $l_{a}\simeq $1 km [c.f. \textit{%
Djuth}, 1984; \textit{Newman et al}., 1998]. At $T_{e}/T_{i}<4$, PDI$_{l}$
is saturated via induced scattering of Langmuir ($l$) waves, piling them up
into `wave condensate' ($k\rightarrow 0$) [e.g., \textit{Zakharov et al}.,
1976]. The condensate is subject to OTSI,\ thereby leading to strong
(cavitating) turbulence and electron acceleration [e.g., \textit{Galeev et al%
}., 1977].

\begin{figure}
\quad \quad \quad \includegraphics[width=18pc]{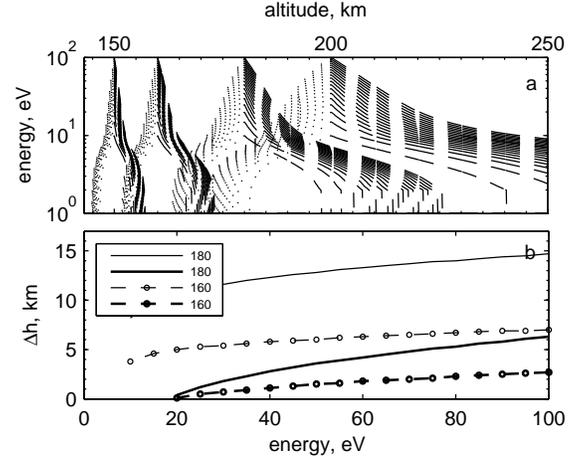}
\caption{ (a) Altitude profiles $\protect%
\varepsilon (\protect\varepsilon _{0},\protect\xi )$ at $\protect\varepsilon %
_{0}=$10, 15, ...100 eV and $h_{0}=$150, ...200 km. (b) Half-widths $\Delta
_{g}\ $(thin lines) and $\Delta _{b}$ (thick) of the green- and blue-line
excitation layers near $h_{c}=$160 (circles) and 180 (solid lines) km.
}
\label{extent}
\end{figure}

At $W_{l}/n_{0}T_{e}<\left( f_{c}/f_{p}\right) ^{2}$, the acceleration
results in a power-law ($\parallel $\textbf{B}$_{0}$) distribution at$%
\mathrm{\,}\varepsilon _{\max }\geq \varepsilon _{\parallel }=\frac{1}{2}%
mu^{2}\geq \varepsilon _{\min }$ [\textit{Galeev et al}., 1983; \textit{Wang
et al}., 1997] 
\begin{equation}
F_{a}^{\parallel }(\varepsilon _{\parallel })\simeq n_{a}(2p_{a}-1)/v_{\min
}\cdot \left( \varepsilon _{\min }/\varepsilon _{\parallel }\right) ^{p_{a}}
\label{1d}
\end{equation}%
where $p_{a}\simeq $0.75-1. The density $n_{a}$ and $\varepsilon _{\min }$
are determined by the wave energy $W_{l}$ trapped by cavitons and the
joining condition with the ambient electron distribution $F_{a}^{\parallel
}(\varepsilon _{\min })=F_{0}(\varepsilon _{\min })$. If $F_{0}$ is a
Maxwellian distribution, this gives $\varepsilon _{\min }^{m}\approx 10T_{e}$
and $n_{a}^{m}\approx 10^{-4}n_{e}$. When background suprathermal ($s$)
electrons of the density $n_{s}$ are present, then $F_{0}(\varepsilon \gg
T_{e})\rightarrow F_{s}(\varepsilon )$ and $\varepsilon _{\min }\simeq
30\left( n_{s}T_{e}/W_{l}\right) ^{2/5}\ $eV [e.g., \textit{Mishin et al.},
2004], yielding $\varepsilon _{\min }\leq 10$ eV at $n_{e}=n_{c}$, $%
\widetilde{W}_{l}\simeq 10^{-3}$, and $n_{s}\leq $10 cm$^{-3}$. In the
ionizing wave, a natural source of the $s$-electrons is ionization by those
accelerated electrons that can propagate from $\xi \sim 0$ to $\xi \sim
L_{\parallel }$ (see Figure~\ref{extent}).

We can now evaluate the excitation and ionization rates. The column 427.8-nm
intensity in Rayleighs (R) is given by 
\begin{equation}
I\approx 10^{-6}A_{b}\int d\xi \int \sigma _{b}(\varepsilon )\Phi
_{a}(\varepsilon \mathbf{,}\xi )d\varepsilon \cdot \lbrack N_{2}(\xi )]
\label{col}
\end{equation}%
Here $\sigma _{b}\ $is the excitation cross section of the $N_{2}^{+}(^{1}N)$
state, $A_{b}\approx $0.19, $\Phi _{a}=\frac{2\varepsilon }{m^{2}}F_{a}$ is
the differential number flux, and $F_{a}(\varepsilon )\simeq n_{a}\frac{%
p_{a}-0.5}{2\pi }v_{\min }^{-3}\left( \varepsilon _{\min }/\varepsilon
\right) ^{p_{a}+1}$ is an isotropic distribution to which the accelerated
distribution $F_{a}^{\parallel }$~(\ref{1d}) is transformed at distances $%
\left\vert \xi \right\vert >v/\nu _{el}$ due to elastic scattering [c.f. 
\textit{Gurevich et al}., 1985].\ 

Integrating eq. (\ref{col}) over the energy range $\varepsilon _{b}\leq
\varepsilon \leq $10$^{2}$ eV at $p_{a}=$0.85 yields the brightness of a $%
\Delta h$-km column $\Delta I(h_{c})\approx 2.5\cdot
10^{-12}n_{a}[N_{2}(h_{c})]\cdot \Delta h$ R near altitude $h_{c}$, given
that $\Delta h\ll H_{n}\simeq $8 km (the atmosphere scale-height). The total
intensity $I$ is defined by the vertical extent of the (excitation) layer $%
\Delta _{b}$, where $\varepsilon (\xi )\geq \varepsilon _{b}$. It can be
evaluated using the \textit{Majeed and Strickland} [1997] loss function $%
L(\varepsilon )=\sum_{j}L_{j}(\varepsilon )$ with $j$ designating $N_{2}$, $%
O_{2}$, and $O$. Outside the acceleration layer, i.e. $\left\vert \xi
\right\vert >l_{a}$, the energy of an electron of the initial energy $%
\varepsilon _{0}$ at a distance $\xi $ from the origination point $h_{0}$ is 
\begin{equation}
\varepsilon (\varepsilon _{0},\xi )\simeq \varepsilon
_{0}-\int_{h_{0}}^{h_{0}+\xi }L(\varepsilon (z))\sqrt{2/\delta
_{e}(\varepsilon (z))}dz  \label{eloss}
\end{equation}

Figure~\ref{extent}a presents the results of calculations of eq. (\ref{eloss}%
) for $\varepsilon _{0}=$10, 15, ... 10$^{2}$ eV and $h_{0}=$150, 160, 180,
and 200 km. The altitude profiles at $\varepsilon \geq $5 eV and hence the
layers of excitation/ionization are nearly symmetric about $h_{0}$ at $h\leq 
$180 km. Panel b shows the half-widths $\Delta _{g}$ and $\Delta _{b}$ of
the green- and blue-line excitation layers about $h_{c}=$180 and 160 km as
function of $\varepsilon _{0}$. The half-width of the ionization layer $%
\Delta _{ion}$ (not shown) is $\approx \Delta _{b}$ at $\varepsilon _{0}>$20
eV. Since $\Delta _{b}<H_{n}$, we can estimate the 427.8-nm intensity at $%
h_{c}=$180$\rightarrow $160 km as $\left. I\right\vert _{h_{c}}\simeq $5$%
\cdot 10^{-12}n_{a}\cdot \lbrack N_{2}(h_{c})]\left\langle \Delta
_{b}(h_{c})\right\rangle \simeq $(0.16$\rightarrow $0.2)$\cdot n_{a}$ R.
Comparing $\left. I\right\vert _{h_{c}}$ with the spatially-averaged
intensities $\widehat{I_{b}}\approx $10$\rightarrow $5 R [P10 Fig.~1] yields 
$\widehat{n}_{a}\simeq $60$\rightarrow $25 cm$^{-3}$. Note that $\widehat{n}%
_{a}\approx n_{a}^{(d)}$ at 180 km, in agreement with a uniform structure,
while spatial averaging underestimates $n_{a}$ inside the $\sim $km-scale
filaments at 160 km.

Calculating the ionization frequency in eq. (\ref{dne}) with $%
F_{a}(\varepsilon )$ gives $\left\langle \nu _{ion}\right\rangle \approx
\kappa _{ion}^{\ast }\cdot \left( \lbrack N_{2}]+\frac{1}{2}%
[O]+0.95[O_{2}]\right) $ s$^{-1}$, where $\kappa _{ion}^{\ast }=\left\langle
v\sigma _{ion}\right\rangle /n_{a}\approx $1.8$\cdot 10^{-8}$ cm$^{3}$s$%
^{-1} $ is the coefficient of ionization of $N_{2}$. The total ionization
rate $q_{a}^{(d)}\sim 10^{4}$ cm$^{-3}$s$^{-1}$ greatly exceeds
recombination losses $\approx 10^{-7}n_{c}^{2}\approx 10^{3}$\ cm$^{-3}$s$%
^{-1}$ (the main ion component at these altitudes is $NO^{+}$). This
justifies the use of eq. (\ref{dne}) for evaluating the artificial plasma
density. Taking an average energy loss per ionization $\sim $20 eV results
in the column dissipation rate $<$0.1 mW/m$^{2}$ or $<$10\% of the 440-MW
Poynting flux, consistent with P10's estimates. 

\begin{figure}
\noindent\includegraphics[width=20pc]{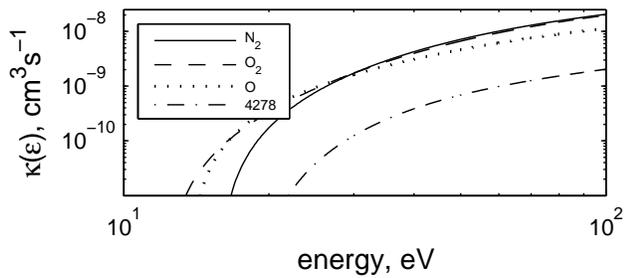}
\caption{ The ionization coefficients of $N_{2}$, $O_{2}$, and $O$ and 
the excitation coefficient of the 
$N_{2}^{+}(^{1}N)$ state vs. $\protect\varepsilon _{\max }$.}
\label{gion}
\end{figure}

As shows Figure~\ref{gion}, the coefficients of ionization and blue-line
excitation by accelerated electrons decrease by a factor of $\sim $2 (10)
between $\varepsilon _{\max }=$10$^{2}$ and 50 (30) eV. The Liouville
theorem predicts $F(\varepsilon _{0}-\Delta \varepsilon (\varepsilon
_{0},\xi ),h_{0}+\xi )=F_{0}(\varepsilon _{0},h_{0})$, where $\Delta
\varepsilon (\varepsilon _{0},\xi )$ is given by the integral in eq. (\ref%
{eloss}). Thus, the gradient scale-length $L_{n}$ of the artificial plasma
is about the distance $\xi _{50}$, defined by the condition $\Delta
\varepsilon (10^{2},\xi _{50})\approx $50 eV. Numerically, we get $\xi
_{50}\approx \Delta _{b}(50)$ or $L_{n}\approx $3$\rightarrow $1.5 km near $%
h_{c}=$180$\rightarrow $160 km and $q_{a}L_{n}/n_{c}\simeq V_{obs}$, as
predicted by eq. (\ref{vdesc}). Note that the artificial plasma density
profiles derived from ionograms indeed have $\sim $1-km gradient
scale-lengths near 150 km [c.f. P10 Fig. 2].

Figure~\ref{layer} shows that the descent slows down below 160 km and
ultimately stops at $h_{\min }\approx $150 km. The presence of IL and bright
green-line emissions indicate that plasma turbulence is still excited and
efficiently accelerates electrons above 4 eV. However, the blue-line
emissions almost vanish [P10], thereby indicating only few accelerated
electrons at $\varepsilon \geq \varepsilon _{b}$. That this is in no way
contradictory follows from the fact that inelastic losses increase tenfold
between 10 and 20 eV. Acceleration stops at $\varepsilon =\varepsilon _{\max
}\ll $100 eV when $\nu _{il}(\varepsilon _{\max })$ exceeds the acceleration
rate $mD_{\parallel }(u_{\max })/8\pi \varepsilon _{\max }$, where $%
D_{\parallel }(u)\approx \frac{\omega _{p}^{2}}{4n_{e}mu}\left\vert
E_{k_{\parallel }}\right\vert ^{2}$ and $k_{\parallel }=\omega _{p}/u$ [%
\textit{Volokitin and Mishin}, 1979]. The critical neutral density is
roughly estimated as $\sim $5$\cdot 10^{11}$ cm$^{-3}$, i.e $N_{n}$ at $\sim 
$150 km. The fact that the artificial plasma stays near $h_{\min }$
indicates that ionization is balanced by recombination or $q_{a}^{\min }\sim
10^{-7}n_{c}^{2}\approx 0.1q_{a}^{(d)}$, which at $n_{a}\sim n_{a}^{(d)}$
corresponds to $\varepsilon _{\max }\approx $30 eV (Figure~\ref{gion}).

A mechanism for generating km-sized filaments below 180 km could be the
thermal self-focusing instability (SFI) near $h_{c}$, resulting in a broad
spectrum of plasma irregularity scale sizes [e.g., \textit{Guzdar et al}.,
1998]. Significantly, $\sim $km-scale plasma irregularities grow initially
but within 10s of seconds thermal self-focusing leads to smaller (10s to
100s meters) scale sizes. During descent, the critical altitude moves
downward by several km within 10 s, thereby precluding further development
of SFI, while the $\sim $km-scale irregularities have sufficient time to
develop. When the descent rate drops, small-scale irregularities can fully
develop and scatter the HF beam, thereby impeding the development of OTSI/PDI%
$_{L}$ and hence ionization. As soon as the artificial plasma decays, SFI
falls away and hence irregularities gradually disappear. Then, the
artificial plasma can be created again. This explains why the artificial
layer ceases and then reappears (Figure~\ref{layer}).

In conclusion, we have shown that the artificial plasma sustaining
interaction with the transmitted HF beam can be created via enhanced
ionization by suprathermal electrons accelerated by Langmuir turbulence near
the critical altitude. As soon as the interaction region is ionized, it
shifts toward the upward-propagating HF beam, thereby creating an ionizing
wavefront, which resembles \textit{Pedersen et al}.'s [2010] descending
artificial ionospheric layers.

\begin{acknowledgments}

{ This research was supported by Air Force Office of Scientific
Research. We thank Chris Fallen for providing the MUIR IL data.}
\end{acknowledgments}

\end{document}